# Room Temperature Intrinsic Ferromagnetism in Epitaxial Manganese Selenide Films in the Monolayer Limit

Dante J. O'Hara,[1,†] Tiancong Zhu,[2,†] Amanda H. Trout,[3,4,†] Adam S. Ahmed,[2] Yunqiu (Kelly) Luo,[2] Choong Hee Lee,[5] Mark R. Brenner,[5,6] Siddharth Rajan,[4,5] Jay Gupta,[2] David W. McComb,[3,4] and Roland K. Kawakami[1,2*]

[1]*Program in Materials Science and Engineering, University of California, Riverside, CA 92521, USA*
[2]*Department of Physics, The Ohio State University, Columbus, OH 43210, USA*
[3]*Center for Electron Microscopy and Analysis, The Ohio State University, Columbus, OH 43212, USA*
[4]*Department of Materials Science and Engineering, The Ohio State University, Columbus, OH 43210, USA*
[5]*Department of Electrical and Computer Engineering, The Ohio State University, Columbus, OH 43210, USA*
[6]*Semiconductor Epitaxy and Analysis Laboratory, The Ohio State University, Columbus, OH 43210, USA*

[†]These authors contributed equally.

*Corresponding Author

    e-mail:    kawakami.15@osu.edu

    Address:    191 W. Woodruff Ave.
    Department of Physics
    The Ohio State University
    Columbus, OH 43210

    Phone:    (614) 292-2515

    Fax:    (614) 292-7557

**ABSTRACT**

Monolayer van der Waals (vdW) magnets provide an exciting opportunity for exploring two-dimensional (2D) magnetism for scientific and technological advances, but the intrinsic ferromagnetism has only been observed at low temperatures. Here, we report the observation of room temperature ferromagnetism in manganese selenide ($MnSe_x$) films grown by molecular beam epitaxy (MBE). Magnetic and structural characterization provides strong evidence that in the monolayer limit, the ferromagnetism originates from a vdW manganese diselenide ($MnSe_2$) monolayer, while for thicker films it could originate from a combination of vdW $MnSe_2$ and/or interfacial magnetism of $\alpha$-MnSe(111). Magnetization measurements of monolayer $MnSe_x$ films on GaSe and $SnSe_2$ epilayers show ferromagnetic ordering with large saturation magnetization of ~ 4 Bohr magnetons per Mn, which is consistent with density functional theory calculations predicting ferromagnetism in monolayer 1T-$MnSe_2$. Growing $MnSe_x$ films on GaSe up to high thickness (~ 40 nm) produces $\alpha$-MnSe(111), and an enhanced magnetic moment (~ 2x) compared to the monolayer $MnSe_x$ samples. Detailed structural characterization by scanning transmission electron microscopy (STEM), scanning tunneling microscopy (STM), and reflection high energy electron diffraction (RHEED) reveal an abrupt and clean interface between GaSe(0001) and $\alpha$-MnSe(111). In particular, the structure measured by STEM is consistent with the presence of a $MnSe_2$ monolayer at the interface. These results hold promise for potential applications in energy efficient information storage and processing.

The study of magnetism in two dimensions (2D) has fascinated physicists for decades, inspiring theoretical studies of phase transitions[1, 2] and topological order[3] as well as their experimental realization in physical systems[4-8]. Recently, intrinsic ferromagnetism has been demonstrated in van der Waals (vdW) crystals in the monolayer limit[9, 10], which creates new opportunities for science and applications related to the potential for highly tunable magnetic properties via electrostatic gating, strain, and proximity effects[11-18]. This is particularly important for spintronics and valleytronics with 2D vdW heterostructures, where the monolayer magnets could provide a route toward low energy magnetization switching and proximity interactions for non-volatile logic.[19-24] However, ferromagnetism in monolayer magnets has so far been limited to low temperatures, below ~ 60 K[9, 10]. In this Letter, we report the observation of room temperature ferromagnetism in manganese selenide ($MnSe_x$) films grown by molecular beam epitaxy (MBE). Magnetic and structural characterization provides strong evidence that in the monolayer limit, the ferromagnetism originates from a vdW manganese diselenide ($MnSe_2$) monolayer, while for thicker films it could originate from a combination of vdW $MnSe_2$ and/or interfacial magnetism of $\alpha$-MnSe(111). This behavior differs of bulk $MnSe_x$ compounds such as $MnSe_2$ (pyrite structure) and $\alpha$-MnSe (rocksalt structure) which are not ferromagnetic[25, 26]. On the other hand, density functional theory (DFT) calculations have predicted the stability of vdW $MnSe_2$ monolayers with 1T structure (Figure 1a) as well as a ferromagnetic ground state with substantial exchange splitting for high Curie temperatures[27, 28]. Our results are consistent with these predictions.

Our investigation consists of material synthesis by MBE, magnetic characterization by superconducting quantum interference device (SQUID) magnetometry, and structural characterization by *in situ* reflection high energy electron diffraction (RHEED), scanning transmission electron microscopy (STEM), scanning tunneling microscopy (STM), atomic force microscopy (AFM), and x-ray diffraction (XRD). To outline our study, we start by growing ~ one monolayer (ML) of manganese selenide ($MnSe_x$) on vdW $SnSe_2$(0001)/GaAs(111) and vdW GaSe(0001)/GaAs(111) substrates. Magnetic measurements by SQUID show ferromagnetism at room temperature, and the similarity of the loop shapes and saturation magnetic moment on two different substrates help eliminate potential artifacts related to substrate interaction. In growing thicker films on GaSe substrates, the RHEED patterns remain streaky up to several tens of nanometers, and XRD scans reveal that the growth converts to rocksalt $\alpha$-MnSe(111). SQUID measurements exhibit room temperature ferromagnetism with larger saturation magnetic moment (~ 2x) than the ~ 1 ML $MnSe_x$, and XRD scans reveal a new peak consistent with vdW transition metal dichalcogenides. Because $\alpha$-MnSe is not ferromagnetic, this provides evidence for ferromagnetic vdW $MnSe_2$ layers forming at the interface of GaSe and $\alpha$-MnSe(111). Detailed structural characterization by STEM, STM, and RHEED reveal an abrupt and clean interface between GaSe(0001) and $\alpha$-MnSe(111). In particular, the structure measured by STEM is consistent with the presence of a $MnSe_2$ monolayer at the interface. This provides strong evidence for the room temperature ferromagnetism originating from vdW MnSe$_2$ monolayers.

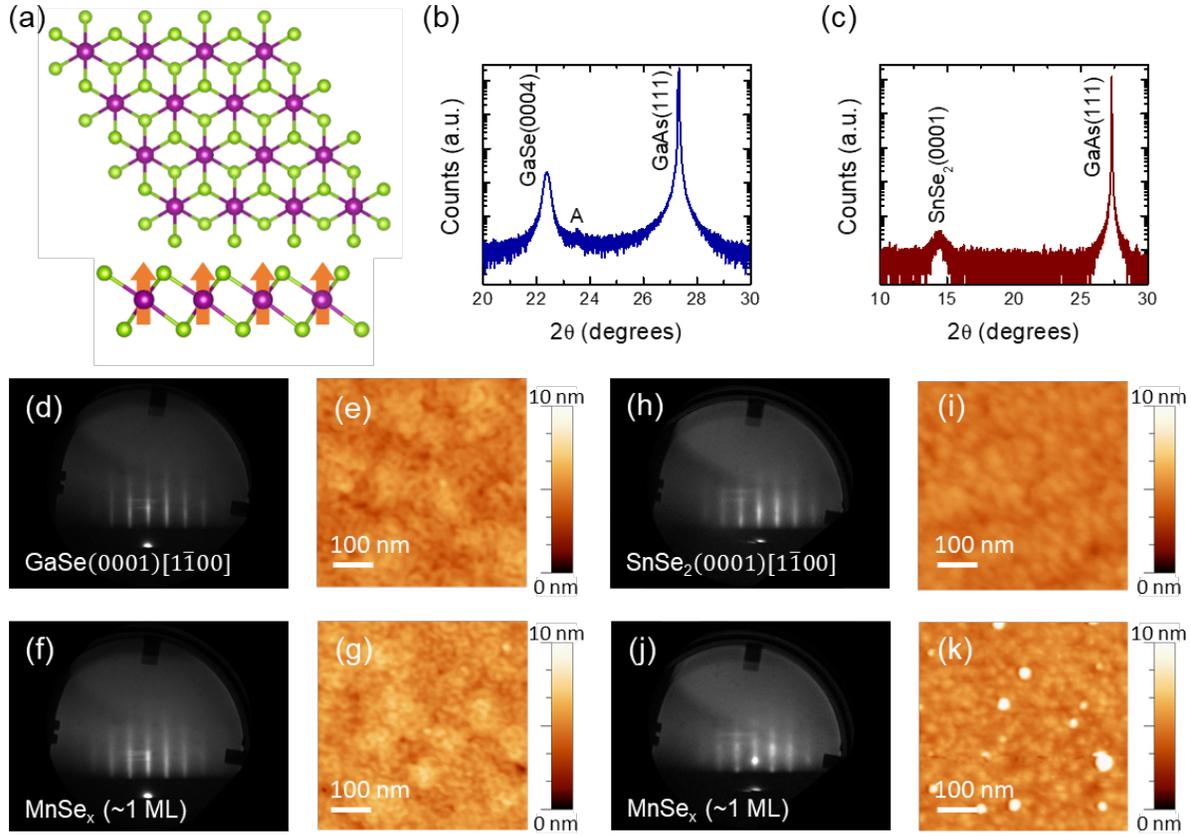

**Fig. 1. MBE growth and structural properties of monolayer MnSe$_x$.** (a) Top and side view of 1T-MnSe$_2$ lattice with purple and green balls representing Mn and Se atoms respectively. Arrows indicate location of magnetic moments. (b, c) θ-2θ XRD scans of ~ 1 ML MnSe$_x$ on 55 nm base layer of GaSe (blue) and on 12 nm base layer of SnSe$_2$ (wine), respectively. (d) RHEED image along the [1$\bar{1}$00] crystallographic axis of GaSe base layer on GaAs(111)B and (e) corresponding AFM image. (f) RHEED image of ~ 1 ML of MnSe$_x$ on GaSe base layer with (g) corresponding AFM image showing smooth morphology. (h) RHEED image along [1$\bar{1}$00] crystallographic axis of SnSe$_2$ on GaAs(111)B and (i) corresponding AFM image. (j) RHEED image of ~ 1 ML MnSe$_x$ on SnSe$_2$ base layer with (k) corresponding AFM image showing island morphology.

The MnSe$_x$ samples are prepared via van der Waals epitaxy[29, 30] in a Veeco GEN930 MBE chamber equipped with a liquid nitrogen cryoshroud and base pressure of 2 × 10$^{-10}$ Torr. We investigate the growth of MnSe$_x$ on two different base layers, GaSe(0001)/GaAs(111)B and SnSe$_2$(0001)/GaAs(111)B. The growth of the GaSe, SnSe$_2$, and MnSe$_x$ layers are performed under a Se overpressure and the growth rate is determined by the Ga, Sn, or Mn flux. Details of the growth are provided in the Supporting Information (SI), sec. 1. For the growth of MnSe$_x$ on the GaSe(0001) surface, we deposit a 55 nm GaSe base layer on GaAs(111)B at 400 °C[31, 32]. Figures 1d and 1e show a streaky RHEED pattern and an AFM image of the GaSe base layer that displays atomically smooth terraces and monolayer steps with a spiral hillocks pattern, respectively. After growth of ~ 1 ML MnSe$_x$ at 400 °C, the RHEED pattern remains streaky (Figure 1f) and the AFM image in Figure 1g shows similar atomically smooth morphology as the GaSe base layer. The RHEED pattern rotates with six-fold rotation symmetry which confirms in-plane epitaxial registry. For characterization by XRD and SQUID, the sample is capped with 5 nm GaSe and amorphous Se to protect the sample from oxidation before being removed from the chamber. The θ-2θ XRD scan for the ~ 1 ML MnSe$_x$ on GaSe (Figure 1b) exhibits two prominent peaks coming from the GaAs(111) substrate and the GaSe base layer (2θ = 22.4°). An additional "A" peak is also observed at 2θ = 23.5°, for which the origin of the peak is not known.

We investigate the magnetic properties of the MnSe$_x$ layers by SQUID magnetometry. Room temperature, out-of-plane magnetization scans reveal a ferromagnetic hysteresis loop for a ~ 1 ML MnSe$_x$ sample on the GaSe base layer (Figure 2a). The inset shows the unprocessed SQUID data which includes diamagnetic and paramagnetic contributions (see SI, sec. 6 for details of background subtraction). The ferromagnetic hysteresis loop exhibits a coercivity of ~ 150 Oe and a saturation magnetic moment of ~ 3.3×10$^{-5}$ emu/cm$^2$ (this represents the total moment, but normalized to the sample area), which corresponds to ~ 4.4 μ$_B$ per Mn.

To rule out the possibility that the ferromagnetism is generated by a substrate-specific artifact, we develop the growth of MnSe$_x$ on another substrate, SnSe$_2$(0001)/GaAs(111)B. For this study, we grow 12nm of SnSe$_2$ on GaAs(111)B at 165 °C[33]. The streaky RHEED pattern (Figure 1h) indicates an atomically smooth surface, which is verified by AFM scans (Figure 1i). After deposition of ~ 1 ML MnSe$_x$ at 165 °C, the RHEED pattern starts becoming spotty indicating 3D growth (Figure 1j) and the corresponding AFM image confirms the 3D growth with island formation as shown in Figure 1k. The RHEED pattern rotates with six-fold rotation symmetry which confirms in-plane epitaxial registry. For XRD and SQUID, an additional protection layer of SnSe$_2$ (12 nm) is deposited and the sample is capped with amorphous Se. The θ-2θ XRD scan (Figure 1c) shows two large peaks from the GaAs(111) substrate and the 12 nm SnSe$_2$ base layer (2θ = 14.5°), and the "A" peak is absent. Room temperature, out-of-plane magnetization scans reveal a ferromagnetic hysteresis loop for a ~ 1 ML MnSe$_x$ sample on the SnSe$_2$ base layer (Figure 2b). The ferromagnetic hysteresis loop exhibits a coercivity of ~ 100 Oe and a saturation magnetic moment of ~ 3.2×10$^{-5}$ emu/cm$^2$ which corresponds to ~ 4.2 μ$_B$/Mn. These values are very similar to that observed in GaSe base layer samples. The absence of the "A" peak while still observing ferromagnetism indicates that any associated structures are not relevant for the magnetic signal.

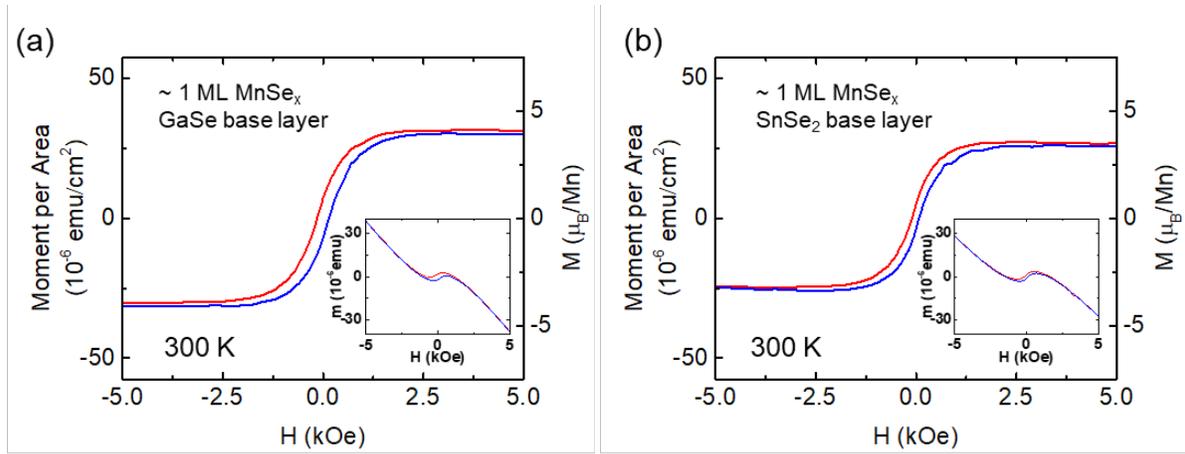

**Fig. 2. Out-of-plane room temperature SQUID magnetometry measurements on ~ 1 ML MnSe$_x$.** (a) Magnetic hysteresis loop of ~ 1 ML MnSe$_x$ on GaSe base layer showing ferromagnetic behavior. Inset: the unprocessed SQUID data prior to background subtraction. (b) Magnetization loop of ~ 1 ML MnSe$_x$ on SnSe$_2$ base layer. Inset: the unprocessed SQUID data prior to background subtraction.

The observation of room temperature ferromagnetism with similar characteristics for both SnSe$_2$ and GaSe base layers implies that the ferromagnetism does not originate from the base layer vdW materials. For example, the formation of certain magnetic compounds such as GaMn[34] in the GaSe samples would provide no explanation for the ferromagnetism observed in the SnSe$_2$ samples. Furthermore, the large magnetic moment (~ 4 $\mu_B$/Mn) cannot be explained by the formation of a dilute magnetic semiconductor (e.g. Ga$_{1-x}$Mn$_x$Se[35] or Sn$_{1-x}$Mn$_x$Se$_2$[36]) due to interdiffusion of Mn into the base layer. Although room temperature ferromagnetism has been observed in Sn$_{1-x}$Mn$_x$Se$_2$ for Mn concentration up to ~ 70%, only a small net saturation magnetic moment of ~ 0.09 $\mu_B$/Mn has been reported[36]. Thus, if the observed ferromagnetism in our samples were due to interdiffusion into the base layer, we would expect characteristics that are different from we observe, and the GaSe and SnSe$_2$ samples should be different from each other. This provides strong evidence that the observed ferromagnetism originates from the deposited MnSe$_x$ monolayers.

To possibly identify the structural composition of MnSe$_x$, we attempt to grow thicker films on GaSe and SnSe$_2$ base layers. For the case of GaSe, the growth of MnSe$_x$ maintains a streaky RHEED pattern through several tens of nanometers. Figures 3a and 3b show the initial RHEED pattern for a 20 nm GaSe base layer and Figures 3c and 3d show RHEED after ~ 40 nm of MnSe$_x$ growth. The θ-2θ XRD scan (Figure 3e) provides an explanation for the persistent streaky pattern. Compared to the XRD scan of the ~ 1 ML MnSe$_x$ on 55 nm GaSe shown in Figure 1b, the main new feature is the emergent of a large peak appearing at 2θ = 28.3°, which indicates the presence of (111)-oriented α-MnSe[37]. α-MnSe is a thermodynamically stable bulk structure[26], which explains the persistence of streaky, crystalline growth out to large thicknesses. We also observe an additional "B" peak at 2θ = 29.4°, which corresponds to a lattice spacing of 6.07 Å (for 2$^{nd}$ order diffraction). This does not correspond to any peak in bulk MnSe$_2$, α-MnSe, or other known Mn-Se compounds [37, 38], but is similar to the layer spacing for vdW layered transition metal dichalcogenides: 6.00 Å for TiSe$_2$, 6.10 Å for VSe$_2$, 6.15 Å for MoS$_2$, 6.16 Å for WS$_2$, 6.46 Å for MoSe$_2$, 6.48 Å for WSe$_2$.[39-43]. In addition to the emergence of new XRD peaks in the thicker MnSe$_x$ films, an enhancement in the magnetization is also observed. As shown in Figure 3f, the magnetic hysteresis loop has similar coercivity (~ 75 Oe) but more than double the saturation magnetic moment per area (~ 7.3×10$^{-5}$ emu/cm$^2$) as compared to the ~ 1 ML MnSe$_x$ sample on GaSe (Figure 2b). Because α-MnSe is not ferromagnetic, the enhanced saturation magnetic moment suggests that the ferromagnetic signal is likely coming from the interface between α-MnSe and GaSe.

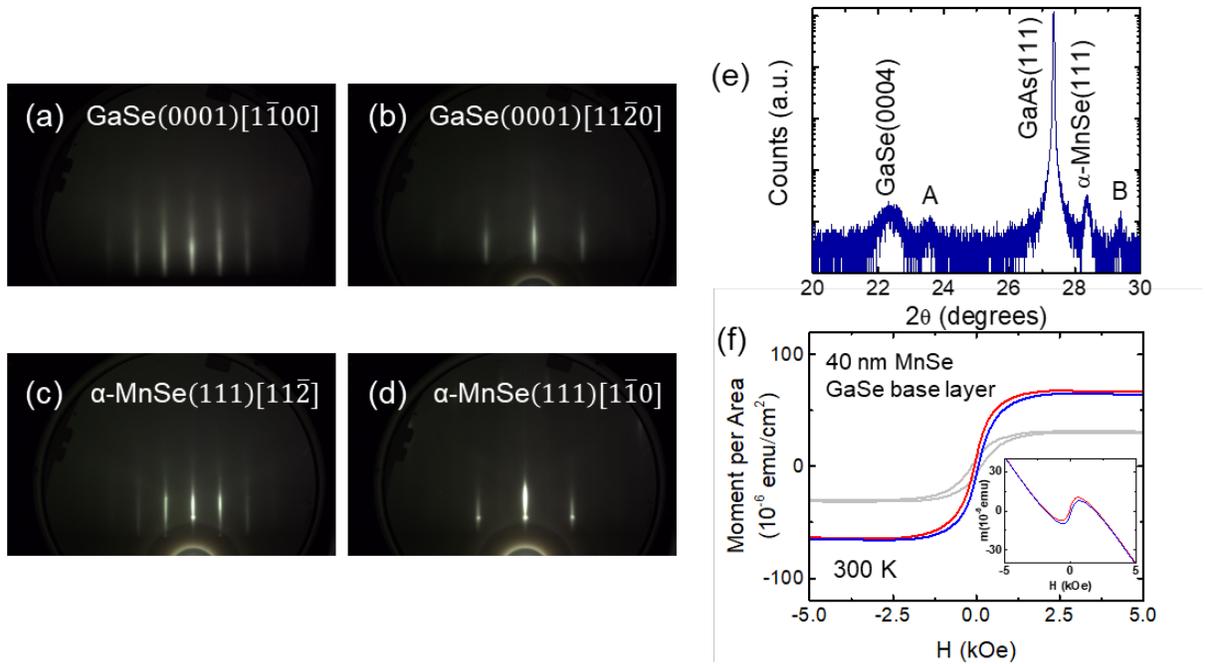

**Fig. 3. Structural and magnetic characterization of thick MnSe$_x$ films.** (a, b) GaSe RHEED images along the [1$\bar{1}$00] and [11$\bar{2}$0] crystallographic axes, respectively, and (c, d) RHEED of 40 nm MnSe$_x$ film epitaxially aligned to the GaSe base layer. (e) θ-2θ XRD scan of a 40 nm MnSe$_x$ film on GaSe showing additional peaks at 28.3° and 29.4°. (f) Magnetic hysteresis loop of a 40 nm MnSe$_x$ film (red and blue) in comparison to ~ 1 ML MnSe$_x$ (grey) showing a larger magnitude of signal.

To better understand the structure at the α-MnSe/GaSe interface, which is likely providing the ferromagnetic ordering, we investigate few-layer MnSe$_x$ films. Figure 4a shows the MnSe$_x$ growth evolution viewing the RHEED images as a function of time during the first ~ 30 seconds (~ 3 ML) of MnSe$_x$ deposition. Linecuts of the [1$\bar{1}$00] RHEED pattern are taken along the white dashed line in the inset of Figure 4a. The red curve shows the linecut for the GaSe base layer, the blue curve shows the linecut after ~ 3 ML of deposition, and the greyscale image shows the evolution of the RHEED linecut during the growth. Notably, the spacing of the RHEED streaks becomes smaller with MnSe$_x$ deposition, which indicates that the in-plane lattice constant increases with MnSe$_x$ deposition. The ratio of the RHEED spacing between GaSe and MnSe is 1.0244, which is very close to the expected ratio of 1.0285 for the bulk in-plane lattice constants of α-MnSe(111) (3.862 Å) and GaSe(0001) (3.755 Å). This confirms what we observed earlier in the XRD scans (Figure 3c) showing the α-MnSe(111) phase. The streakiness and six-fold rotational symmetry of the RHEED patterns also suggests that we have epitaxial alignment between each material ($[11\bar{2}]_{MnSe} \parallel [1\bar{1}00]_{GaSe}$ and $[1\bar{1}0]_{MnSe} \parallel [11\bar{2}0]_{GaSe}$). Figure 4b shows the in-plane lattice constant of the MnSe$_x$ film normalized by the GaSe in-plane lattice constant ($a_{film}/a_{GaSe}$) as a function of time and thickness. The structural transition from the GaSe base layer is abrupt and transitions to the bulk α-MnSe lattice constant within one monolayer of MnSe$_x$ deposition. The formation of α-MnSe(111) at the MnSe$_x$/GaSe interface is further confirmed by STM. The surface structure, lattice constant (3.90 Å), and band gap (~ 3.39 eV) measured with dI/dV spectroscopy on a ~ 3 ML MnSe$_x$ sample (Figure 4d) are in good agreement with that of bulk α-MnSe [44], which is distinct from the same measurement performed on GaSe base layer (Figure 4c).

The cross-sectional STEM high angle annular dark field (HAADF) images, Figure 4e, demonstrate high quality growth of GaSe[32] on GaAs with the epitaxial relationship of $[111]_{GaAs} \parallel [0002]_{GaSe}$ (see SI, sec. 7 for details). We observe the γ-GaSe polytype in two orientations that are related by a 30° in-plane rotation, which is very useful for the STEM imaging study. Because the α-MnSe overlayer is registered with the GaSe lattice, the $[1\bar{1}00]_{GaSe}$ and $[11\bar{2}0]_{GaSe}$ viewing directions can be imaged simultaneously without any tilting of the specimen. Figure 5a shows HAADF images that contains two different orientations of α-MnSe, including high magnification images (Figure 4e(i) and (ii)) that clearly show the rotated atomic lattices. Both have the relationship $[111]_{α-MnSe} \parallel [0002]_{GaSe}$, however (i) is oriented along <110> and (ii) along <112>. The former also confirms the similarity between α-MnSe along <110> and 1T-MnSe$_2$ along $[11\bar{2}0]$ as discussed below. The structure in both orientations is consistent with the presence of a MnSe$_2$ monolayer at the interface and it should be noted that if 1T-MnSe$_2$ along $[11\bar{2}0]$ also undergoes a 30° in-plane rotation, the atoms align vertically similar to the GaSe $[1\bar{1}00]$ direction shown in the schematic in Figure 4e. Although, we have no direct evidence from imaging or compositional mapping (see SI, sec. 8) for the presence of this phase. In all cases, the interface between the GaSe and MnSe layers is abrupt with no evidence for segregation or contaminants.

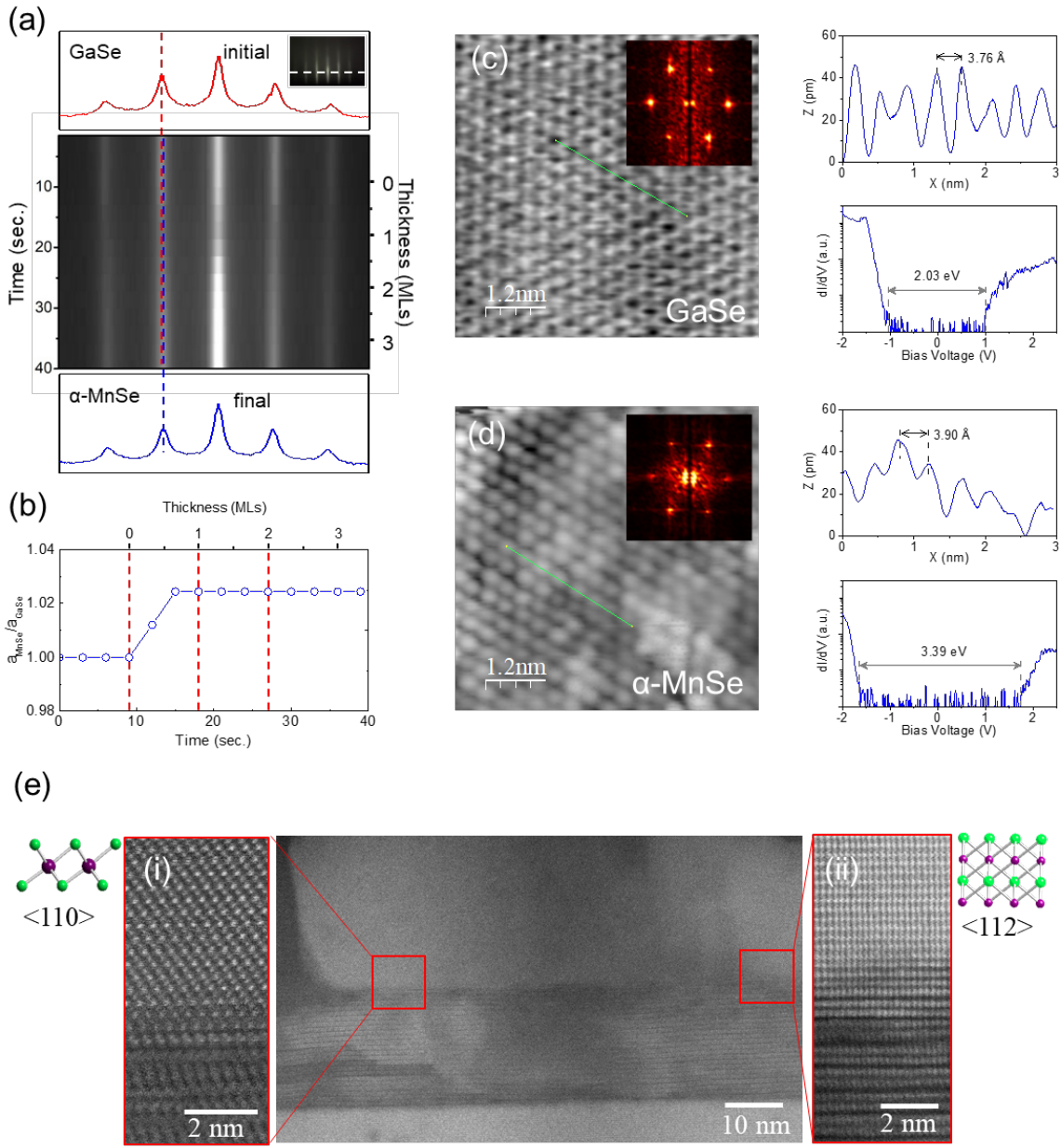

**Fig. 4. RHEED line profile and high-resolution imaging of α-MnSe/GaSe interface.** (a) RHEED line profile of MBE growth evolution from GaSe base layer to ~ 3.5 layers of MnSe$_x$ (inset showing region of GaSe RHEED image that the line profile is taken from). (b) Lattice constant ratio as a function of thickness showing abrupt change from GaSe to α-MnSe(111) lattice within ~ 1 ML of MnSe$_x$ deposition. (c) STM topography image, Fourier transform, line profile and dI/dV spectrum on a GaSe base layer sample. The image is taken with an etched PtIr tip, with V = -2.0 V and I = 0.2 nA. (d) Same set of data taken on another sample with ~ 3 ML MnSe$_x$ sample grown on GaSe. (e) High resolution STEM HAADF imaging of α-MnSe(111) and GaSe with different orientations. (i) α-MnSe(111) viewed along <110> and (ii) shows α-MnSe viewed along <112>. Both images are collected from the areas indicated by the boxes in the middle image. There is no tilting of the specimen between imaging the two areas. These samples are grown at 300 °C.

The clean, sharp and crystalline interface between GaSe and α-MnSe(111) observed in STEM indicates that the monolayer MnSe$_x$ should have similar structure as the α-MnSe(111). This brings a key insight on the origin of the observed ferromagnetism in monolayer MnSe$_x$ (Figure 2). It is important to notice that a single monolayer of α-MnSe(111) is virtually identical to a monolayer of vdW MnSe$_2$ with 1T structure. This is depicted in Figures 5a and 5b which show the top view, side view, and nearest-neighbor coordination diagrams (insets) for the α-MnSe(111) and vdW 1T-MnSe$_2$ lattices, respectively. The similarity of the structures is evident in the top view and nearest-neighbor coordination diagrams. Most importantly, the side view of the lattices shows clearly that a single monolayer of α-MnSe(111), which is the Se-Mn-Se atomic trilayer highlighted by the dashed lines in Figure 5a, is equivalent to the 1T-MnSe$_2$ shown in the side view of Figure 5b. Furthermore, DFT calculations show that the 1T structure is thermodynamically stable for monolayer MnSe$_2$ and is ferromagnetic close to room temperature[27, 28]. Their predicted magnetic moment (3.0-3.7 μ$_B$/Mn) is consistent with our experimental observation in Figure 2. In addition, as discussed earlier in Figure 4, the smooth evolution of the RHEED pattern from GaSe to α–MnSe without intermediate patterns, the rapid transition to α-MnSe(111) as confirmed by STM, and the STEM images provide strong evidence that the first monolayer forms in the 1T-MnSe$_2$ structure. Based on these considerations, we conclude that the room temperature ferromagnetism in monolayer MnSe$_x$ originates from magnetic ordering of vdW 1T-MnSe$_2$.

It is also important to discuss the ferromagnetism in the thick MnSe$_x$ samples which exhibit slightly larger magnetic moment per area than in the ML. We consider two possible mechanisms that could contribute to the magnetic signal. One is the stabilization of one or more vdW MnSe$_2$ layer at the interface. This would be consistent with the experimental results from STEM and XRD ("B" peak). The other possibility is that the surface of α-MnSe(111) could exhibit interfacial ferromagnetism, such as that observed in Cr$_2$O$_3$.[45] It would be very interesting to investigate these two possibilities in future studies. Nevertheless, due to the structural similarities between α-MnSe(111) and 1T-MnSe$_2$, there is no distinction between these two mechanisms in the monolayer limit.

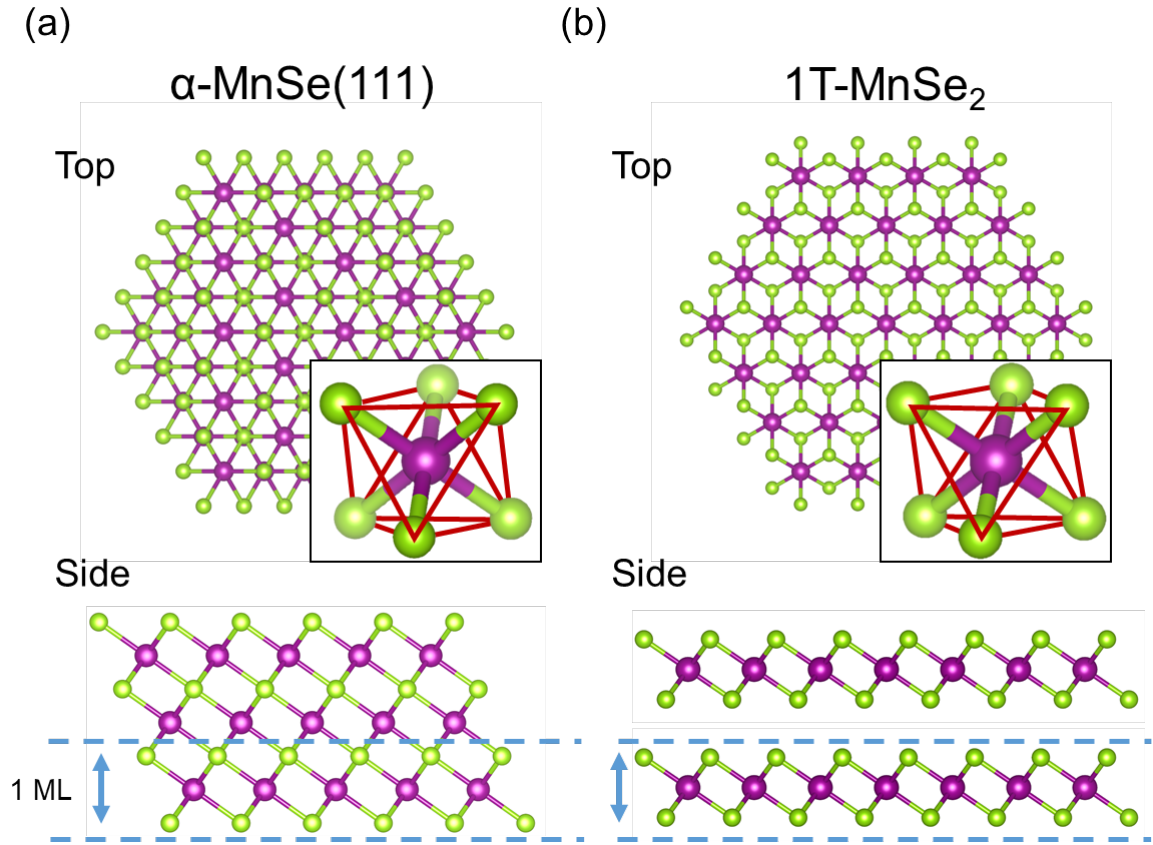

**Fig. 5. Crystal structure diagrams of α-MnSe(111) and monolayer 1T-MnSe$_2$.** (a) Ball-and-stick model of α-MnSe(111) hexagonal lattice (top view) with inset showing octahedral coordination. The Mn atom is purple and the Se atom is green. The side view shows that 1 ML of α-MnSe(111) is equivalent to 1 ML 1T-MnSe$_2$. (b) Ball-and-stick model of 1T-MnSe$_2$ hexagonal lattice (top view) with inset showing octahedral coordination. The side view shows that 1T-MnSe$_2$ is equivalent to 1 ML α-MnSe(111).

In conclusion, we have observed room temperature ferromagnetism in epitaxial manganese selenide films grown by MBE. In the monolayer limit, we attribute the magnetic signal to intrinsic ferromagnetism of a vdW manganese diselenide (MnSe$_2$) monolayer, while for thicker films it could originate from a combination of vdW MnSe$_2$ and/or interfacial magnetism of α-MnSe(111). This enables the integration of room temperature ferromagnetism into 2D layered vdW heterostructures in a variety of ways. For monolayers, the vdW MnSe$_2$ could be grown on appropriate vdW surfaces (either exfoliated or epitaxial layers) and capped with other vdW materials to produce the isolated MnSe$_2$ vdW layers. This could be used for studies of vertical magnetic tunneling junctions, magnetic proximity effect, and gate tunable magnetism. For thick α-MnSe(111) structures, its insulating character could allow it be used as a gate dielectric for spin field-effect structures and gate tunable magnetic proximity effect. In addition, direct measurement of the magnetism and atomic scale magnetic ordering by spin-polarized STM could help realize the full potential of 2D magnets for spintronics and valleytronics.

## ASSOCIATED CONTENT

**Supporting Information**

Details of material synthesis, structural and magnetic characterization methods, a comparison of growth at 300 °C and 400 °C, in-plane magnetic hysteresis loops, data from additional samples, details of SQUID magnetization loop background subtraction, additional cross-sectional STEM images, and elemental mapping by energy dispersive x-ray spectroscopy (EDS).

## AUTHOR INFORMATION

**Corresponding Author**

*E-mail: kawakami.15@osu.edu

**Author Contributions**

D. J. O. and R. K. K. conceived the project. D. J. O., T. Z., C. H. L., M. R. B., S. R., and R. K. K. participated in the material synthesis. A. H. T. and D. W. M. performed the STEM imaging and analysis. T.Z., J. G, and R. K. K. performed the STM imaging and analysis. D. J. O., T. Z., A. S. A., Y. K. L., and R. K. K. performed the SQUID and AFM measurements. All authors contributed to writing the manuscript.

## ACKNOWLEDGEMENTS

We thank Camelia Selcu, Sriram Krishnamoorthy, Steven Tjung, Jacob Repicky, Takahiro Takeuchi, and Igor Pinchuk for technical assistance. This project was supported by the US Department of Energy (DE-SC0018172) and National Science Foundation MRI program for the MBE system (DMR-1429143). D. J. O. acknowledges the GEM National Consortium Ph.D. Fellowship. We acknowledge support from The Ohio State University Materials Research Seed Grant Program, funded by the Center for Emergent Materials, an NSF-MRSEC (DMR-1420451), the Center for Exploration of Novel Complex Materials, and the Institute for Materials Research.

**Supporting Information for:**

**Room Temperature Intrinsic Ferromagnetism in Epitaxial Manganese Selenide Films in the Monolayer Limit**

Dante J. O'Hara,[1,†] Tiancong Zhu,[2,†] Amanda H. Trout,[3,4,†] Adam S. Ahmed,[2] Yunqiu (Kelly) Luo,[2] Choong Hee Lee,[5] Mark R. Brenner,[5,6] Siddharth Rajan,[4,5] Jay Gupta,[2] David W. McComb,[3,4] and Roland K. Kawakami[1,2*]

[1]*Program in Materials Science and Engineering, University of California, Riverside, CA 92521, USA*
[2]*Department of Physics, The Ohio State University, Columbus, OH 43210, USA*
[3]*Center for Electron Microscopy and Analysis, The Ohio State University, Columbus, OH 43212, USA*
[4]*Department of Materials Science and Engineering, The Ohio State University, Columbus, OH 43210, USA*
[5]*Department of Electrical and Computer Engineering, The Ohio State University, Columbus, OH 43210, USA*
[6]*Semiconductor Epitaxy and Analysis Laboratory, The Ohio State University, Columbus, OH 43210, USA*

[†]These authors contributed equally.

*Corresponding Author

    e-mail:    kawakami.15@osu.edu

    Address:    191 W. Woodruff Ave.
                   Department of Physics
                   The Ohio State University
                   Columbus, OH  43210

    Phone:    (614) 292-2515

    Fax:    (614) 292-7557

## 1. Details of material growth

Manganese selenide layers are grown by MBE in a Veeco GEN930 ultra-high vacuum (UHV) chamber with a base pressure of $2\times10^{-10}$ Torr. Epi-ready semi-insulating un-doped GaAs(111)B substrates (AXT, single-side polished, 0.5 mm thick, ± 0.5° miscut, $1.4\times10^{8}$ Ω-cm resistivity) are indium-bonded to unpolished Si backing wafers, loaded into the chamber and annealed at 400 °C for 15 minutes under UHV conditions ($1\times10^{-9}$ Torr) to remove any surface impurities. The substrates are then loaded into the growth chamber onto a continuous azimuthal rotation (CAR) manipulator and annealed at 600 °C for 10 minutes to remove the native oxide layer. The substrate temperature is measured using a thermocouple that is attached to the CAR substrate heater. The substrate is then terminated at 600 °C with Se (BEP $\sim2\times10^{-6}$ Torr) for 20 minutes. A commercial valved cracking source (Veeco) is used to evaporate Se (United Mineral & Chemical Corporation, 99.9999%) with a bulk zone temperature of 290 °C and cracking zone temperature of 950 °C. The samples are monitored via *in situ* RHEED to obtain real-time feedback about the crystalline quality of the growth. The sample is then cooled down to 400 °C or 165 °C under an Se background (Se shutter remaining open) to initiate the GaSe or $SnSe_2$ growth, respectively. We use standard Knudson-style thermal effusion cells to evaporate Ga (United Mineral & Chemical Corporation, 99.99999%), Sn (Alfa Aesar, 99.9999%), and Mn (Alfa Aesar, 99.98%) with typical cell temperatures of 1000 °C, 1100 °C and 800 °C, respectively. The MBE growth of GaSe, $SnSe_2$, and $MnSe_x$ is performed in an adsorption-limited growth regime[1], where Se is the volatile species. The BEP flux ratios are as follows: Se/Ga ~ 100, Se/Sn ~ 40, and Se/Mn ~ 60 for Se-rich conditions, where the excess Se re-evaporates. The beam fluxes are measured using a nude ion gauge with a tungsten filament positioned at the sample growth position and the corresponding deposition rate is calibrated based on film thicknesses determined by x-ray reflectometry (XRR). These growths are then followed by opening the Mn shutter (with Se shutter open always) for approximately 9 seconds to grow ~ one monolayer of $MnSe_x$. Subsequently, an overlayer of GaSe or $SnSe_2$ is grown and the sample is cooled to room temperature. The sample is then capped with amorphous Se to protect it from oxidation and the sample is then removed from the chamber.

For STM measurements, we utilize conducting GaAs substrates that are mounted onto Omicron flagship style sample holders, which is loaded onto a Veeco uni-block adapter for the MBE growth. After growth, the sample and Omicron holder are transferred to the STM without air exposure via UHV suitcase. Due to the different sample mounting, the actual substrate temperature for the STM samples is different than the actual substrate temperature for the standard sample mounting using the Si backing wafers for equivalent thermocouple temperatures. Therefore, we rely on the RHEED patterns and their evolution during $MnSe_x$ deposition to ensure that the growth occurs within the temperature window from 300 – 400 °C (details of growths at these temperatures is discussed in Supporting Information, Section 2). The conducting substrates used for the STM samples are epi-ready, semiconducting, *n*-type, Si-doped

GaAs(111)B (AXT, single-side polished, 0.5 mm thick, ± 0.5° miscut, (0.8~4)×10$^{18}$ cm$^{-3}$ carrier concentration).

## 2. Growth of thick MnSe$_x$ films at 300 °C and 400 °C

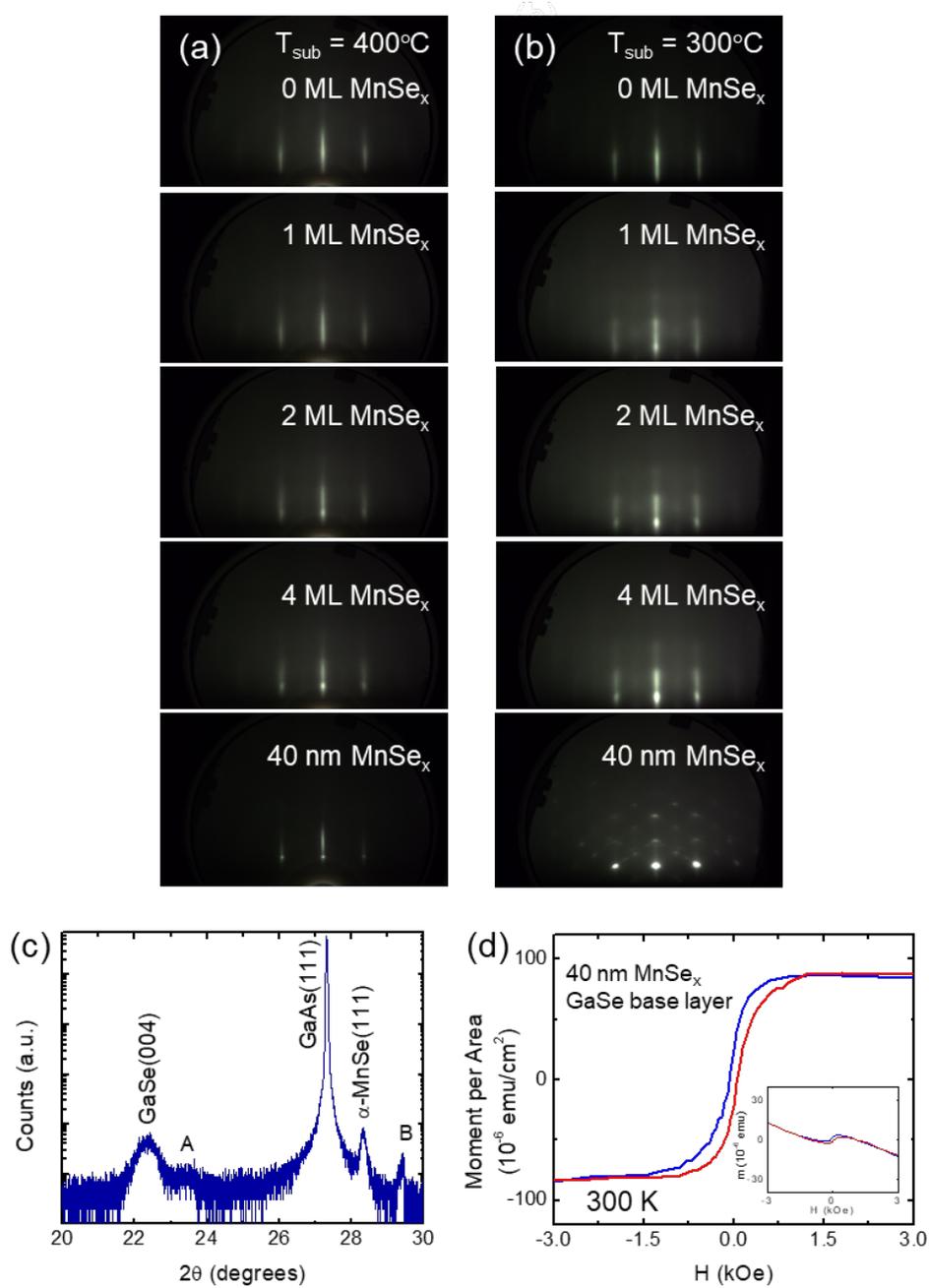

**Fig. S1. Growth of thick MnSe$_x$ films at 300 °C and 400 °C.** (a) RHEED growth evolution from GaSe base layer to 40 nm MnSe$_x$ grown at 400 °C compared to (b) MnSe$_x$ growth at 300 °C. (c) θ-2θ XRD scan of 40 nm MnSe$_x$ grown at 300 °C on GaSe base layer showing same peaks as 40 nm MnSe$_x$ grown at 400 °C. (d) Room temperature out-of-plane magnetic hysteresis loop of 40 nm MnSe$_x$ grown at 300 °C showing similar ferromagnetism as 40 nm MnSe$_x$ grown at 400 °C (Inset: Raw data prior to background subtraction).

Due to the important insights obtained about the magnetic and structural properties of $MnSe_x$ from the thick $MnSe_x$ films grown at 400 °C (observed in Figure 3), we explored growth of $MnSe_x$ at different temperatures. For $MnSe_x$ growth at 300 °C, a GaSe(0001) base layer is initially grown at 400 °C and the sample is then cooled to 300 °C (with all shutters closed). Mn and Se are then deposited at 300 °C on the GaSe base layer. Figures S1a and S1b show a comparison of RHEED patterns during $MnSe_x$ growth for substrate temperatures of 400 °C and 300 °C, respectively, from the first few monolayers up to ~40 nm thickness. The RHEED pattern in the initial stages of growth (~ few MLs of $MnSe_x$ deposition) remains streaky for both temperatures, suggesting similar interfacial structure. However, with increasing thickness, for growth at 300 °C the streaks evolve into 3D-like spots, suggesting a rougher morphology. Nevertheless, the cross-sectional STEM images obtained for the 300 °C growth show that the interface and film have very good atomic-scale ordering (Figure 4). Figures S1c and S1d show XRD scans and magnetic properties for 300 °C growth, and their characteristics are similar to the 400 °C growth (Figures 3e and 3f). The θ-2θ XRD scans show the same peaks that were observed in Figure 3e and the room temperature out-of-plane magnetic hysteresis loop has similar shape and magnitude to Figure 3f.

## 3. Structural and magnetic characterization methods

X-ray diffraction measurements of the MBE-grown GaSe, $SnSe_2$, and $MnSe_x$ films are performed in a Bruker, D8 Discover system equipped with Cu-Kα 1.54 Å wavelength x-ray source. Tapping mode atomic force microscopy (AFM) is performed in a Bruker Icon 3 system to measure the morphology of the films. Image analysis is performed with the *WSxM* software[2]. Cross-sections are prepared for STEM imaging using a focused ion beam (FIB) instrument (FEI Helios) with final polishing using low energy argon ions in a Fischione Nanomill. STEM imaging is performed in a FEI Titan 60-300 aberration corrected STEM operating at 300 kV. Magnetic properties are measured via SQUID magnetometry (Quantum Design, MPMS XL) using the reciprocating sample option (RSO). M(H) loops are measured using the max slope position and linear regression fitting parameters to eliminate centering errors at zero moment. *VESTA* software[3] is used to generate crystal structure schematics of monolayer 1T-$MnSe_2$ and rocksalt α-MnSe(111). STM measurements are performed with a CreaTec LT-STM/AFM system operating at 4.3 K and chamber base pressure ~ $2\times10^{-10}$ Torr. An etched PtIr tip is used for the STM measurements on the samples. Prior to measuring each $MnSe_x$ sample, we ensure that the tip is free of anomalies in its electronic structure by performing calibration dI/dV measurements on a clean Au(111) surface. Image analysis is performed with the *WSxM* software[2].

## 4. Room temperature out-of-plane and in-plane magnetic hysteresis loops

Figures S2a and S2b show out-of-plane and in-plane magnetization loops for 40 nm MnSe$_x$ on GaSe base layers and ~ 1 ML MnSe$_x$ on SnSe$_2$ base layers, respectively. For the GaSe base layer sample (Figure S2a), the in-plane and out-of-plane loops show similar shapes, so there is no indication of a strong uniaxial magnetic anisotropy. For the SnSe$_2$ base layer sample (Figure S2b), the in-plane and out-of-plane loops have different shapes. The larger saturation field of the in-plane loop suggests an out-of-plane magnetic easy axis. Further studies based on ferromagnetic resonance (FMR) would provide a more direct characterization of the magnetic anisotropy.

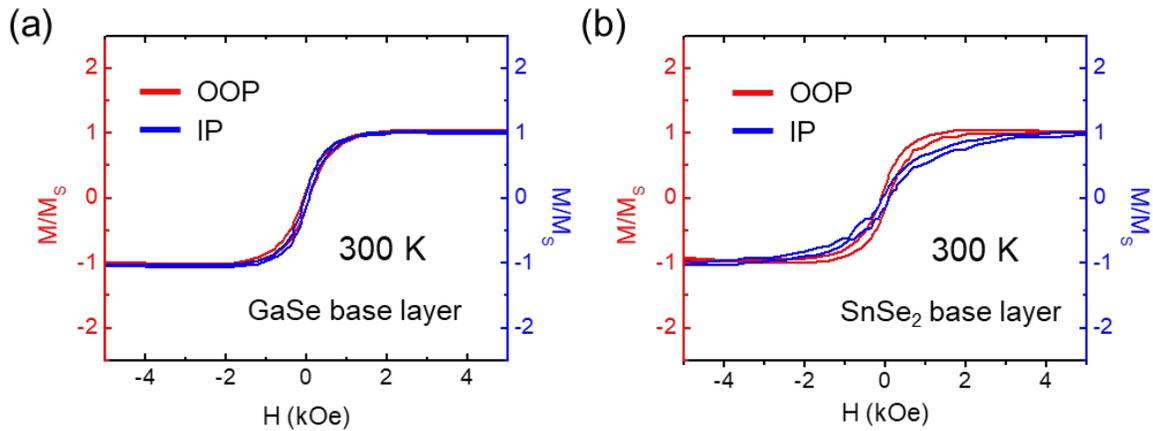

**Fig. S2. Room temperature out-of-plane and in-plane magnetic hysteresis loops.** (a) In-plane and out-of-plane magnetic hysteresis loops of 40 nm MnSe$_x$ on GaSe base layer. (b) In-plane and out-of-plane magnetic hysteresis loops of ~ 1 ML MnSe$_x$ on SnSe$_2$ base layer.

## 5. Data from additional samples

Figure S3 shows out-of-plane magnetic hysteresis loops measured at room temperature for additional samples grown under similar conditions as those in the main text. The loop shapes are generally similar, and the sample-to-sample variations appear mostly related to the saturation magnetic moment. Figure S3a is from ~ 1 ML MnSe$_x$ on a 55 nm GaSe base layer and exhibits a saturation magnetization of 4.2 $\mu_B$/Mn Figure S3b is from ~ 1 ML MnSe$_x$ on a 12 nm SnSe$_2$ base layer and exhibits a saturation magnetization of 3.6 $\mu_B$/Mn Figure S3c is from 40 nm MnSe$_x$ on a 55 nm GaSe base layer and exhibits a saturation moment per area of $9.1 \times 10^{-5}$ emu/cm$^2$.

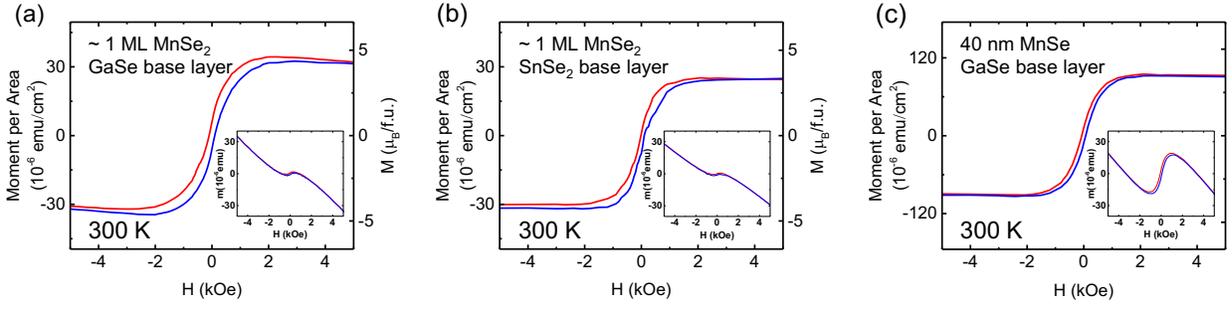

**Fig. S3. Data from additional samples.** (a) Magnetic hysteresis loop of ~ 1 ML MnSe$_x$ on GaSe base layer showing ferromagnetic behavior. Inset: the unprocessed SQUID data prior to background subtraction. (b) Magnetic hysteresis loop of ~ 1 ML MnSe$_x$ on SnSe$_2$ base layer. Inset: the unprocessed SQUID data prior to background subtraction. (c) Magnetic hysteresis loop of 40 nm MnSe$_x$ on GaSe base layer. Inset: the unprocessed SQUID data prior to background subtraction.

## 6. SQUID magnetization loop background subtraction

The SQUID magnetization loops have diamagnetic and paramagnetic background contributions. The diamagnetic background comes from the GaAs substrate. The paramagnetic background may come from trace unintentional Mn doping of the substrate and/or base layer. We utilize the following procedure for background subtraction.

Figure S4a shows the raw SQUID data for the sample in Figure 4a. For the background subtraction, we select a cutoff field, $H_{cut}$, that bounds the range of the ferromagnetic loop. Later, we will discuss the dependence of the fit on the choice of $H_{cut}$. To demonstrate the fitting method, we proceed with $H_{cut} = 2$ kOe (vertical black dashed lines in Fig. S4a) and fit the data outside of $H_{cut}$ with the following equation

$$m(H) = A_{dia} \cdot H + A_{para} \cdot B_{5/2}\left(\frac{\frac{5}{2}g^*\mu_B}{k_B T} \cdot H\right) + sign(H) \cdot m_{Sat} \qquad (S1)$$

where $A_{dia}$, $A_{para}$, $g^*$, and $m_{sat}$ are fitting parameters. The first term is a diamagnetic background that is linear in $H$. The second term is a paramagnetic background described by a Brillouin function with $J = 5/2$ for Mn, and $g^*$ is an effective g-factor as observed in dilute magnetic semiconductors[4]. The last term is the saturated ferromagnetic magnetization, which adds as a positive/negative offset ($m_{sat}$) depending on the direction of the applied magnetic field. After the fit is completed, the first two terms of equation S1 (i.e. the diamagnetic and paramagnetic contributions) are subtracted from the raw data to yield the ferromagnetic hysteresis loop. Figures S4c and S4d shows the raw data, diamagnetic component, paramagnetic component, and ferromagnetic component over the field range of +/-5 kOe (Fig. S4c) and the full field range (Fig. S4d). This yields a value for $m_{sat}$ of 4.4 μ$_B$/Mn.

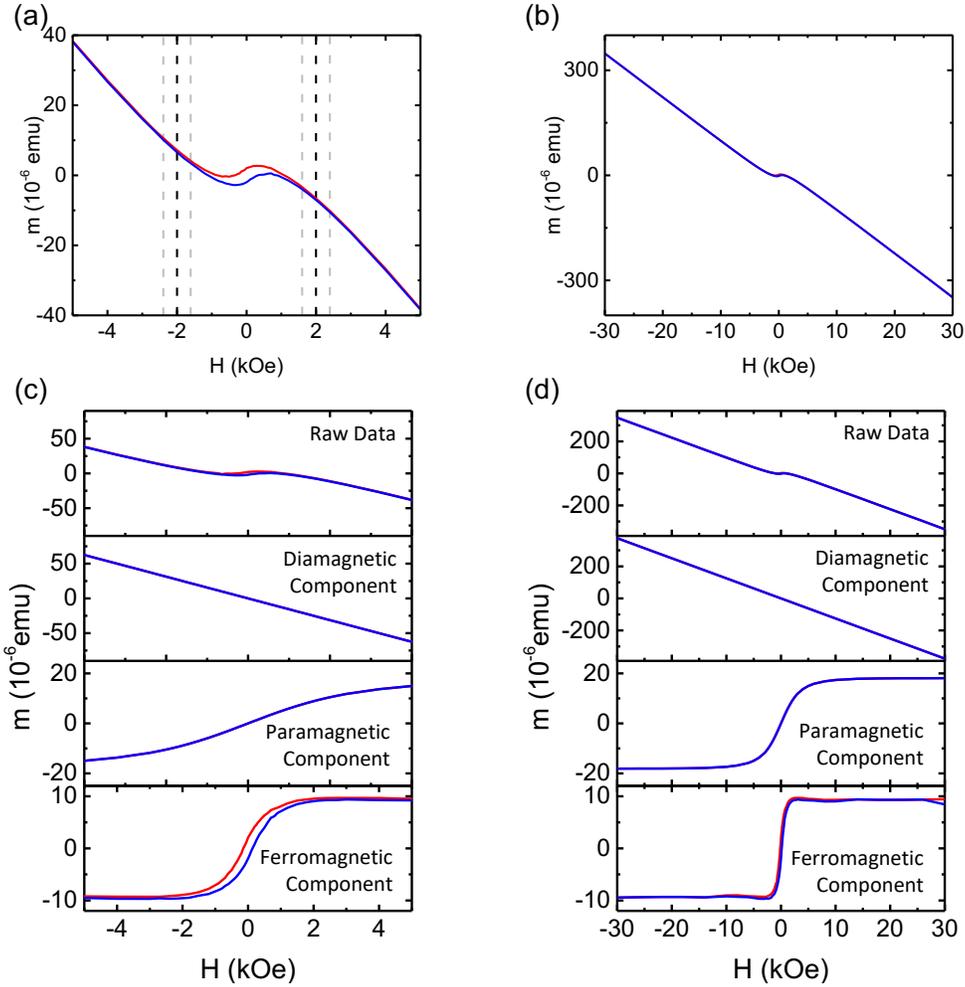

**Fig. S4. Magnetization loop background subtraction.** (a) Raw data for sample in Figure 2a of main text (~1 ML MnSe$_x$ on GaSe base layer). Dashed black lines represent $H_{cut}$, while the grey dashed lines show the range bounded by $H_{cut}^{min}$ and $H_{cut}^{max}$. (b) Raw data plotted over the full field range. (c) The raw data, diamagnetic component, paramagnetic component, and ferromagnetic component plotted from –5 kOe to +5 kOe field range. (d) The raw data, diamagnetic component, paramagnetic component, and ferromagnetic component plotted over the full field range.

Because the fitting depends on the choice of $H_{cut}$, we specify an acceptable range for $H_{cut}^{min}$ to $H_{cut}^{max}$ (see grey dashed lines at $H_{cut}^{min}$ = 1600 Oe and $H_{cut}^{max}$ = 2400 Oe in Figure S4a). Repeating the fit for these values of $H_{cut}$ yields a range of $m_{sat}$ between 3.8 – 5.0 μ$_B$/Mn. This procedure was performed on the seven samples presented in the main text and Supporting Information and the results are summarized in the following table:

| Description | Figure number | $H_{cut}$ | $m_{sat}$ | $H_{cut}^{min} - H_{cut}^{max}$ | $m_{sat}^{min} - m_{sat}^{max}$ |
|---|---|---|---|---|---|
| ~1 ML MnSe$_x$ on GaSe | 2a | 2.00 kOe | 4.4 $\mu_B$/Mn | 1.6 – 2.4 kOe | 3.8 – 5.0 $\mu_B$/Mn |
| ~1 ML MnSe$_x$ on SnSe$_2$ | 2b | 1.85 kOe | 4.2 $\mu_B$/Mn | 1.5 – 2.2 kOe | 3.9 – 4.4 $\mu_B$/Mn |
| 40 nm MnSe$_x$ on GaSe | 3f | 2.10 kOe | 7.3 × 10$^{-5}$ emu/cm$^2$ | 1.7 – 2.5 kOe | 6.5 – 7.5 × 10$^{-5}$ emu/cm$^2$ |
| 40 nm MnSe$_x$ on GaSe | S1d | 1.65 kOe | 9.2×10$^{-5}$ emu/cm$^2$ | 1.3 – 2.0 kOe | 8.8 – 9.3 × 10$^{-5}$ emu/cm$^2$ |
| ~1 ML MnSe$_x$ on GaSe | S3a | 1.75 kOe | 4.2 $\mu_B$/Mn | 1.5 – 2.0 kOe | 3.5 – 4.6 $\mu_B$/Mn |
| ~1 ML MnSe$_x$ on SnSe$_2$ | S3b | 1.35 kOe | 3.6 $\mu_B$/Mn | 1.0 – 1.7 kOe | 3.1 – 4.4 $\mu_B$/Mn |
| 40 nm MnSe$_x$ on GaSe | S3c | 1.65 kOe | 9.1 × 10$^{-5}$ emu/cm$^2$ | 1.5 – 1.8 kOe | 8.3 – 9.1 × 10$^{-5}$ emu/cm$^2$ |

## 7. Cross-sectional STEM imaging

The cross-sectional STEM high angle annular dark field (HAADF) images demonstrate high quality growth of GaSe on GaAs(111). The GaSe layer contains occasional defects such as indicated by the red arrow in Figure S5, as well as the presence of multiple polytypes. As previously reported, GaSe can grow in different polytypes on GaN with β-GaSe and ε-GaSe as the primary polytypes present, while a third polytype was also grown but was unable to be identified[5]. These different polytypes are related by in plane rotations of a GaSe layer, resulting in a different stacking sequences. For GaSe grown on GaAs (Figure S6), we observe the γ-GaSe polytype in two orientations that are related by a 30° in-plane rotation. The [1$\bar{1}$00] viewing direction of the GaSe layer makes identification of other polytypes difficult because they appear identical along this orientation.

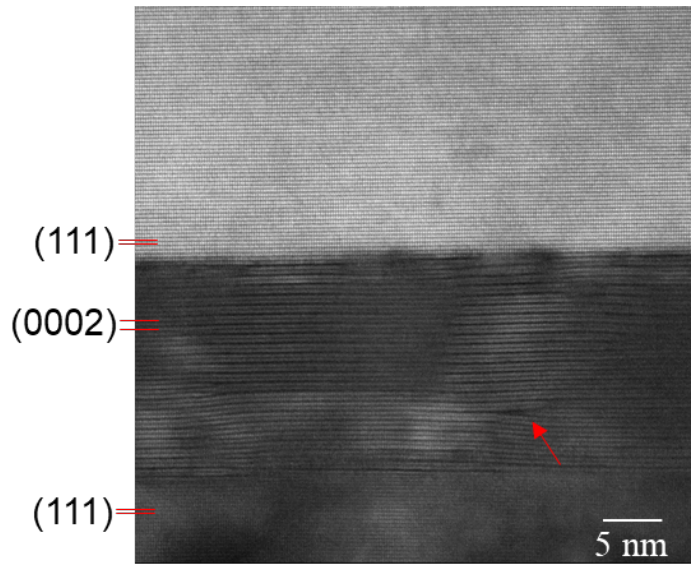

**Fig. S5. HAADF image of α-MnSe/GaSe/GaAs.** The GaSe layer consists of lattice planes and occasional defects as indicated by the red arrow.

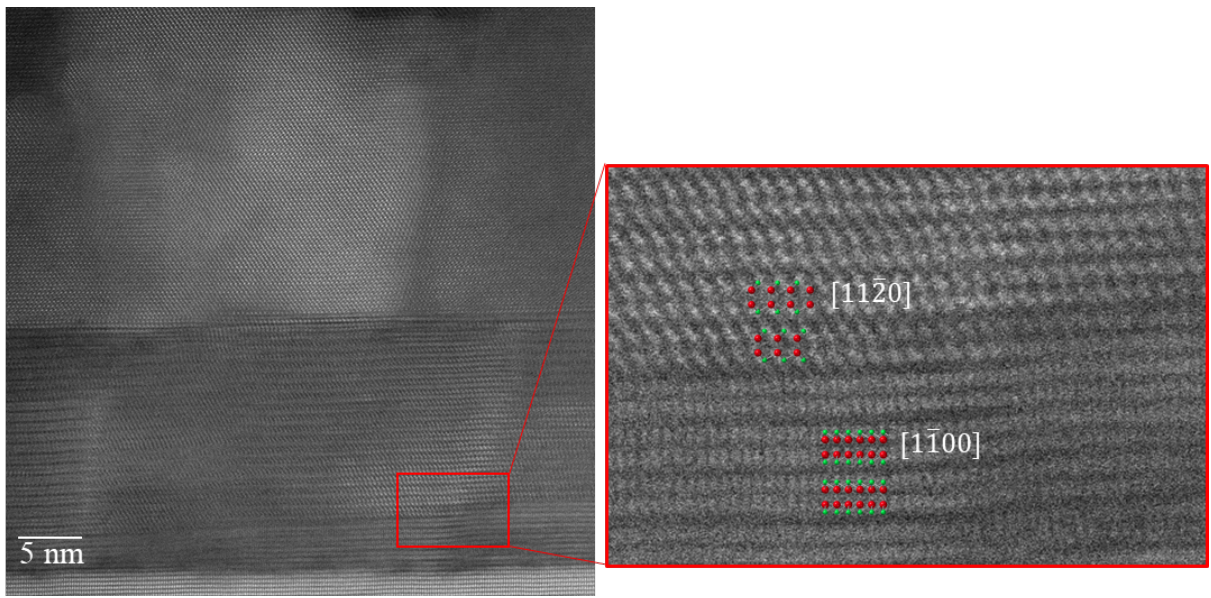

**Fig. S6. HAADF image of α-MnSe/GaSe/GaAs showing GaSe polytypes.** (Left) Cross-sectional image of a typical α-MnSe/GaSe/GaAs(111) heterostructure in a region exhibiting multiple polytypes. (Right) Higher magnification image showing in-plane rotation in the GaSe layers. Because the α-MnSe is registered with the GaSe lattice, this results in rotated α-MnSe grains in the overlayer. Overlaid schematic of γ-GaSe structure show the two orientations caused by a 30° rotation. The viewing direction for the two orientations are $[11\bar{2}0]$ for the top schematic and $[1\bar{1}00]$ for the bottom schematic.

## 8. Energy dispersive X-ray spectroscopy (EDS)

The depth profile of the chemical composition is measured by EDS on an FEI Probe Corrected Titan 80-300 S/TEM at 300kV with approximately 300 pA of current. The standard-less Cliff-Lorimer method is used for quantification. The MnSe$_x$ layer shows a nearly 1:1 ratio of the atomic % for Mn:Se stoichiometry. The GaSe layer shows a nearly 1:1 ratio of the atomic % for Ga:Se stoichiometry.

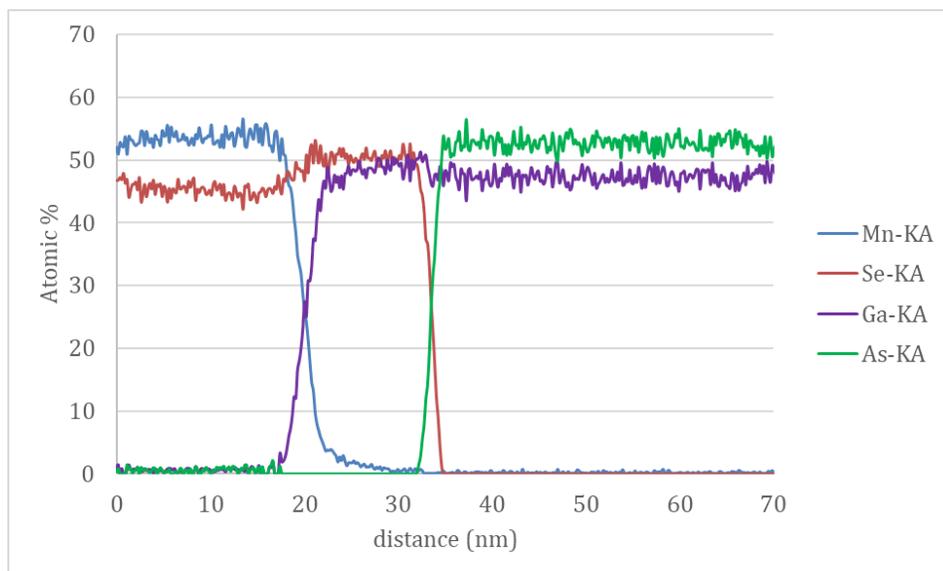

**Fig. S7.** EDS compositional mapping of α-MnSe/GaSe/GaAs(111).